\documentclass[final]{aipproc}
\layoutstyle{6x9}
\pdfoutput=1

\begin{document}
\title{Unification for Yukawas and its implications}

\author{Enkhbat Tsedenbaljir}{
  address={The International Centre for Theoretical Physics, 34014 Trieste, Italy}
}

\begin{abstract}
The supersymmetric finite threshold effects are studied in the
presence of non minimal soft terms that can correct the problematic
mass ratios of the light generations in the minimal $SU(5)$ GUT. We
show that with large soft $A$--terms, one can achieve simple
unification for lighter generations without additional Higgs
multiplet, while having sfermions lighter than $1$ TeV. The presence
of such large $A$--terms will distort the sfermion mass spectrum upon
running from GUT scale down to the electroweak scale making it distinct
from the universal SUSY breaking scenarios, especially in the first two generations. The
implications of these splittings are studied in $K$ and $D$ meson oscillations and
in rare processes $D^+\rightarrow \pi^+\nu\bar{\nu}$ and
$K^+\rightarrow \pi^+\nu\bar{\nu}$, and in the latter case the effect
is found to be important.

\end{abstract}

\maketitle

\section{The SUSY threshold effects on the quark masses}
The appeal for the supersymmetry (SUSY) is enforced by the fact that
the gauge couplings unify nicely while the $b$ and  the $\tau$ Yukawas
unify at a reasonably good level. These successes are not extended to
the lighter generations: $m_d/m_s$ is an order of magnitude larger
than $m_e/m_{\mu}$ instead of being equal if unified values are
assumed at the grand unification (GUT) scale. The most pursued
solutions to this so called problem of ``wrong GUT ratios''
include the introduction of new Higgs multiplets such as ${\bf 45}$ in
$SU(5)$ \cite{Georgi:1979df} or higher dimensional Planck mass suppressed operators.

Another option is due to the
idea that the unification for the Yukawas of the lighter generations is obscured by
the SUSY finite threshold effects at the electroweak scale
\cite{Buchmuller:1982ye}. In particular, the
gluino--squark loop induced correction to the mass of the $i$th generation down--type quark  is given by
\begin{eqnarray}\label{eq2}
\left(\delta m_d\right)_i&\simeq&-\frac{2\alpha_s}{3\pi}(m^d_{LR})_im_{\tilde{g}}
I\left( m^2_{\tilde{Q}_i},m^2_{\tilde{d}^c_i},m_{\tilde{g}}^2\right),\\
m^d_{LR}&=&A_dv_d-Y_d\mu v_u.
\end{eqnarray}
The most of the studies on these corrections have been in the limit of
large $\tan\beta$ \cite{Hall:1993gn}. In the case of universal soft terms one has
$m^d_{LR}=m_d(A_0-\mu \tan\beta)$, which leads to universal change
\begin{eqnarray}\label{eq3}\frac{\delta m_{d_i}}{m_{d_i}}\simeq\frac{\alpha_s}{3\pi}\frac{(A_0-\mu \tan\beta)m_{\tilde{g}}}{ m^2_{\tilde{d}_i}}.\end{eqnarray}
Here we approximated the loop integral as $I\left(
  m^2_{\tilde{Q}_i},m^2_{\tilde{d}^c_i},m_{\tilde{g}}^2\right)\simeq1/(2m_{\tilde{d}_i}^2)$. This
expression for the leading SUSY QCD effect clearly shows that if the soft masses are universal so
are the induced relative changes in the quark masses. On the other hand, the needed change for
the $d$ and $s$ quark masses are very different. If the down--type quark Yukawa couplings
are set to that of the charged leptons at the GUT scale, the value of the down quark mass before the correction is less than its
experimental value by $\sim1.5$ MeV while that of the $s$--quark is greater by $\sim0.16$
GeV at $M_{SUSY}=1$ TeV. Thus it seems inevitable to depart from a universal assumption for
the soft terms. Indeed the scans over the universal soft parameters by
several studies (See for example \cite{DiazCruz:2000mn}) have found no solution for the wrong GUT
ratios.

Looking at the formula we see that there are two option available: choose either (i) different squark
masses or (ii) non minimal $A$--terms that are not
proportional to the Yukawa couplings. The first option requires $m^2_{\tilde{d}}/m^2_{\tilde{s}}\simeq 2.3$ for
inducing $\delta m_d=-3.5$ MeV and $\delta m_s=-0.16$ GeV which will unlikely survive the
constraint from the neutral Kaon mixing unless the squarks are quite heavy, in the range of  TeVs.
This leaves us with the second option if one wants have sfermions spectrum within the
reach of the LHC.


To induce the needed corrections using the $A$--terms, the numerical values
for them have to be much larger than the most SUSY breaking scenarios such as mSUGRA. 
There is a constraint on the $A$--terms from the stability of the vacuum
\begin{eqnarray}\label{eq4}
\left(A_d\right)_{ij}\leq 1.75\sqrt{\frac{1}{3}\left(m^2_{\tilde{Q}_i}+m^2_{\tilde{d}^c_j}+m^2_{H_d}+\mu^2\right)}.
\end{eqnarray}
When satisfied it guarantees the stability of our vacuum at
cosmological time scale \cite{Kusenko:1996jn} and allows much larger parameter space than
the severe constraint \cite{Casas:1996de} of absolute stability. 

\section{Large $A$--terms for unification and their implications}

The presence of large $A$--terms will have several observable
consequences due to their effects on the RGE running of the soft
masses. Here we display several examples of large $A$--terms for
various choices of $\tan\beta$, which lead
to the SUSY finite correctiions that give the correct low energy $d$ and
$s$ quark masses. The corrections to the charged lepton masses are
small at least by a factor   $3\alpha^\prime/(8\alpha_S)\simeq0.03$ compared to the
quarks and found to be at the level of less than a few percent. For
this reason, the initial values for the Yukawa couplings of the
down--type quarks are set to that of the charged
leptons at the GUT scale. We assume the soft terms satisfy $SU(5)$ boundary condition at the GUT scale along with
the gauge and the Yukawa couplings. The soft masses are chosen to have
universal forms and the $A$--term of the up sector is given by
$A_u=a_0Y_u$. On the other hand, we let the $A$--terms of the down
sector have large values that are not proportional to their respective
Yukawa couplings. To avoid the FCNC constraints, we keep them in
diagonal form in the basis where the corresponding Yukawa matrix is
diagonal:
\begin{eqnarray}\label{eq5}
\left(A_5\right)_{ij}\equiv \left(A_{\ell}\right)_{ij}=\left(A_{d}\right)_{ij}= a_i\delta_{ij}\neq a_0Y_{d_i}.
\end{eqnarray}  
In Ref.~\cite{DiazCruz:2000mn}, the large $A$--terms were used also
for generating the Cabibbo mixings in addition to correcting the wrong
GUT ratio. This led to heavy sfermions
$\geq4.4$ TeV mainly to avoid the constraint from $\mu\to
e\gamma$. Since our primary concern is not a 
possible origin of flavor structure but the unified common values
for the Yukawas, we abondon that choice in favor
of the form given in Eq.~(\ref{eq5}). 
\begin{table}
\begin{tabular}
{lllll}\hline
$\tan\beta$   & 5      & 10     & 15     &20
\\ \hline
$\mu$ (GeV)       & 500  & 550   & 580     &850
 \\ \hline
 $A_d$ (GeV) & 3.5    & 6.4     & 9.2     &16.6 \\ \hline
 $A_s$  (GeV) &  -280    & -460     & -760     &-900   \\ \hline
 $A_b$  (GeV) &  -900    & -950     & -800     &-228     \\ \hline
\hline
$\delta m_{d}$ (MeV)      & 1.50      & 1.43     &
1.55     &1.69  \\ \hline
$\delta m_{s}$ (GeV)      & -0.170      & -0.167     &
-0.158     &-0.156   \\ \hline
$\delta m_{b}$ (GeV)      & -0.730      & -0.732     &
-0.697     &-1.0     \\ \hline
\end{tabular}
\caption{The choices for the $\mu$--term and relevant soft trilinear
  $A$--terms at low energy and the induced change to the down--type quark masses.}
\label{tabledelm}
\end{table}
Here we choose the following values for the soft masses $\{m_{1/2},
m^2_{\tilde{10}}, m^2_{\tilde{5}}\}=\{-210(-230), 560(580),
523(542)\}$ GeV for $\tan\beta=5, 10(15,20)$ at the GUT scale.
The corresponding $A$--terms that induce the needed corrections to the down--type
quarks are given in Table~\ref{tabledelm}. In the numerical calculation, we have included the subleading
neutralino and chargino corrections. See Ref.~\cite{Enkhbat:2009jt}
for details. Our choices for the soft parameters
indeed give the finite contributions that lead to quark masses in agreement with their
experimentally determined values. For $\tan\beta=20$,  we also show,
as an example, the case where the correction to the $b$--quark mass is
somewhat larger than what one needs. Such cases could be easily
compatible with experiment if,upon embedding to a concrete GUT model, the right--handed neutrino effect on the $\tau$ Yukawa
RGE running is included.

Large $A$--terms in the first two generation will split the
squark masses during the RGE running from the GUT scale to
the electroweak scale. This could lead to excessive neutral meson
mixings. We have calculated the induced Kaon and $D$--meson mass
differences and they, as shown in Table~\ref{table5},  have been found to be at a safe level. Here the
leading contributions are from the chargino and the gluino boxes for
Kaon and the $D$--meson respectively. As for $B$ and $B_s$ mesons, the
induced mass differences are expected to be small at low $\tan\beta$
values we have chosen.

\begin{table}
\begin{tabular}
{lllll} \hline
$\tan\beta$              & 5      & 10     & 15     &20  \\ \hline
$\Delta M_{D}\times10^{14} $ GeV$^{-1}$      & 1.44$\times$ 10$^{-2}$      & 0.120     &
0.59     &1.42     \\ \hline
$\Delta M_{K}\times10^{15} $ GeV$^{-1}$      & 1.59$\times$ 10$^{-2}$      & 0.105     &
0.706     &1.55      \\ \hline
\end{tabular}
\caption{The mass splittings in $K$ and $D$ meson systems due to the
  SUSY effects.}
\label{table5}
\end{table}

We have looked into rare decays $D^+\rightarrow
\pi^+\bar{\nu}\nu$ and $K^+\rightarrow \pi^+\bar{\nu}\nu$ since they do
not have large long distance contributions. In the $D^+$ meson case the
effect is, although $\sim10^4$ times larger than the SM prediction,
still far from the reach of BESIII. On the other hand, we find that the
effect from the squark mass splitting could be quite important for
$K^+$ and in some cases could be somewhat larger than the experimental
results. The numerical results are given in  Table~\ref{table6}. This
contribution could be checked when
the experimental precision is refined in future experiments \cite{Anelli:2005ju}.

\section{Conclusions}

In this talk I have presented the solution to the wrong GUT ratios
through finite SUSY corrections and several of its experimental consequences. With large $A$--terms one can achieve
the minimal Yukawa unification with sub TeV sfermion masses. Through
RGE running the squark masses will split due to the large $A$--terms making the spectrum distinct
from the widely studied universal scenarios. The charm and
strange meson oscilations and their rare decays are studied. In
particular, $K^+\rightarrow \pi^+\bar{\nu}\nu$ process has been found
to have a substantial SUSY contribution.

\begin{table}
\begin{tabular}
{lllll} \hline
$\tan\beta$              & 5     & 10     & 15     &20  \\ \hline
$Br(D^+\rightarrow \pi^+\bar{\nu}\nu)\times10^{11} $       & 0.0259      & 0.128     &
8.39     & 0.699    \\ \hline
$Br(K^+\rightarrow \pi^+\bar{\nu}\nu)\times10^{11} $      & 0.141      & 3.73     &
37.8     & 19.3    \\ \hline
\end{tabular}
\caption{The branching ratios for processes $D^+\rightarrow
  \pi^+\bar{\nu}\nu$ and $K^+\rightarrow \pi^+\bar{\nu}\nu$.}
\label{table6}
\end{table}

\begin{theacknowledgments}
The author would like to thank the organizers of the SUSY 09, in particular
professor Pran Nath, for the invitation to a nice and stimulating conference.
\end{theacknowledgments}

\end{document}